\documentstyle[12pt,cite]{article}
              
\title{Covariance Systems on the Projective Line.}
\author{N. Pinamonti and M. Toller   \\ 
Department of Physics of the University and I.N.F.N.  \\
I-38050 Trento, Italy}

\begin{document} 
\maketitle                             
                 
\begin{abstract}
We study the positive-operator-valued measures on the projective real line covariant with respect to the projective group, assuming that the energy is a positive operator. This problem is similar to the more complicated problem of finding the positive-operator-valued measures on the compactified Minkowski space-time covariant with respect to the conformal group. It also describes a time-of-arrival observable for a free particle covariant with respect to linear canonical transformations. 

\bigskip  \bigskip

\end{abstract}

\newpage

\section{Introduction.}  

In the classical textbooks of quantum mechanics, the observables are described by self-adjoint operators or by the corresponding spectral measures. Several authors remarked that some measuring instruments define observables which require a more refined mathematical description in terms of positive-operator-valued measures (POVMs) \cite{Holevo,Davies,BLM,BGL}. If $\hat {\cal M}$ is the the topological space of the possible results of the measurement, for every Borel subset $I \subset \hat {\cal M}$ one gives a positive operator $\tau(I)$ and one assumes that all these operators form a normalized countably additive POVM on the space $\hat {\cal M}$. If the normalized vector $\psi$ belonging to the Hilbert space $\cal H$ represents a physical state, $(\psi, \tau(I) \psi)$ is the probability that the result of the measurement belongs to $I$.

If the theory has a symmetry group $\cal G$ acting on the Hilbert space $\cal H$ by means of its unitary representation $g \to U(g)$ and on the space $\hat {\cal M}$ by means of its (not necessarily linear) representation $g \to \Lambda(g)$, it is natural to impose the covariance condition
\begin{equation}  \label{Covariance}
U(g) \tau(I) U(g)^{\dagger} = \tau(\Lambda(g)I).
\end{equation}
In this case the POVM $\tau$ and the unitary representation $U$ form a {\it covariance system}. In many physical situations the covariance property permits the explicit construction of relevant classes of POVMs.

In the recent years, these ideas have been applied to deal with many problems which could not be treated by means of the usual formalism. For instance, the time observable, which cannot be described by a self-adjoint operator \cite{Pauli} has been treated in a completely satisfactory way \cite{BGL2,Giannitrapani,AOPRU,MLP} in terms of a POVM defined on the time axis and covariant with respect to the time-translation group.  In a similar way one can treat the four space-time coordinates of an event in terms of  a POVM defined on the Minkowski space-time covariant with respect to the Poincar\'e group \cite {Toller}.

It has been shown \cite{JR,JR2} that the form of the operators describing the coordinates of an event can be determined in a natural way in a theory symmetric with respect to the conformal group, as the theory which describes non-interacting photons. Since these operators cannot be self-adjoint \cite{Wightman}, it seems useful to reformulate the problem in terms of a POVM on the compactified Minkowski space-time, covariant with respect to the conformal group. This problem is treated in a forthcoming article \cite{PT}, but it involves rather complicated calculations which obscure the physical and mathematical arguments. For this reason, in the present article we study a much simpler problem in which the group ${\cal G} = SU(2, 2)$ (a four-fold covering of the conformal group) is replaced by ${\cal G} = SU(1, 1)$ and the compactified Minkowski space-time $\hat {\cal M}$ is replaced by the compactified time axis (indicated by the same symbol). Practically all the relevant features of the original problem appear in a simplified form in this model. 

The group $SU(1, 1)$ is composed of all the $2 \times 2$ complex matrices $g$ with the properties
\begin{equation} \label{Def}
g^+ \beta g = \beta, \qquad \det g = 1. 
\end{equation} 
In the standard definition of $SU(1, 1)$ one puts
\begin{equation}
\beta = \left( \begin{array}{cc}
1 & 0 \\ 0 & -1
\end{array} \right),
\end{equation}
but for our purposes it is more convenient to perform a change of basis and to adopt the definition
\begin{equation}
\beta = \left( \begin{array}{cc}
0 & 1 \\ -1 & 0
\end{array} \right).
\end{equation}
Then the condition $\det g = 1$ is equivalent to the condition 
\begin{equation}
g^T \beta g = \beta
\end{equation} 
and, comparing with eq.\ (\ref{Def}), we see that $g$ must be real. In conclusion, our group $\cal G$, isomorphic to $SU(1, 1)$,  is just $SL(2, R)$, which coincides with the symplectic group $Sp(2, R)$. It is also isomorpic to a double covering of the proper orthochronous three-dimensional Lorentz group $SO^{\uparrow}(1, 2)$.

We put
\begin{equation}
g = \left( \begin{array}{cc}
a & b \\ c & d
\end{array} \right), \qquad ad - bc = 1,
\end{equation}
where $a, b, c, d$ are real numbers. If $d \neq 0$, we can always consider the decomposition
\begin{equation} \label{Decomp}
g = \left( \begin{array}{cc}
1 & t \\ 0 & 1
\end{array} \right)
\left( \begin{array}{cc}
d^{-1} & 0 \\ 0 & d
\end{array} \right)
\left( \begin{array}{cc}
1 & 0 \\ v & 1
\end{array} \right),
\end{equation}
where
\begin{equation}
t = \frac b d, \qquad v = \frac c d.
\end{equation}

We consider the subgroup ${\cal S} \subset SL(2, R)$ containing all the real matrices of the form
\begin{equation}
s = \left( \begin{array}{cc}
d^{-1} & 0 \\ 0 & d
\end{array} \right)
\left( \begin{array}{cc}
1 & 0 \\ u & 1
\end{array} \right), \qquad d \neq 0,
\end{equation}
and the homogeneous space $\hat{\cal M} = SL(2, R) / {\cal S}$. We indicate by  ${\cal M} \subset \hat{\cal M}$ the space of the cosets composed of matrices with $d \neq 0$. From the decomposition (\ref{Decomp}) we see that these cosets contain  representative elements of the form
\begin{equation}
\left( \begin{array}{cc}
1 & t \\ 0 & 1
\end{array} \right)
\end{equation}
and we can identify $\cal M$ with the real time axis. All the matrices with $d = 0$ form a single coset, interpreted as the point at infinity of the real projective line $\hat{\cal M}$.

In order to see how $SL(2, R)$ operates on $\hat{\cal M}$, we write
\begin{equation}
\left( \begin{array}{cc}
a & b \\ c & d
\end{array} \right)
\left( \begin{array}{cc}
1 & t \\ 0 & 1
\end{array} \right) = 
\left( \begin{array}{cc}
1 & t' \\ 0 & 1
\end{array} \right) s, \qquad s \in {\cal S},
\end{equation}
where
\begin{equation} \label{ProTrasf}
t' = \Lambda(g) t = \frac {at + b} {ct + d}.
\end{equation}
This equation can be extended in a natural way to the case in which $t$ or $t'$ are infinite and it describes the projective transformations of $\hat{\cal M}$. We see that the first two matrices in the right hand side of eq.\ (\ref{Decomp}) represent, respectively, time translations and time dilatations.

Given a covariant POVM on the time axis, one can define a time operator by means of the formula
\begin{equation} \label{Time}
T = \int_{\hat{\cal M}} t \, d\tau.
\end{equation}
A covariance property of the kind
\begin{equation} 
U^{\dagger}(g) T U(G)  = \Lambda(g) T
\end{equation}
follows from the covariance property (\ref{Covariance}) of the POVM $\tau$ only if $\Lambda(g)$ is linear, namely if $g$ belongs to the affine subgroup of $\cal G$, generated by the time translations and the time dilatations. A possible operator which has these properties is given by
\begin{equation} \label{Time2}
T = E^{-1/2} D E^{-1/2},
\end{equation}
where $E$ is the energy operator and $D$ is the generator of the time dilatations defined in Section 3.  This formula is similar to the one suggested in refs.\ \cite{JR,JR2} for the operators which describe the space-time coordinates.  One may ask under which conditions eqs.\ (\ref{Time}) and (\ref{Time2}) define the same operator.
 
The identity $SL(2, R) = Sp(2, R)$ suggests a physical interpretation in terms of linear canonical transformations of a free particle in one dimension. We consider first the classical (non-quantum) case, we assume that the mass is $m = 1$ and we indicate by $q(t)$ and $p(t)$ the canonical variables. The {\it time-of-arrival} $t$ at which the particle reaches the origin $q = 0$ is given by 
\begin{equation}
t = - \frac {q(0)} {p(0)}.
\end{equation} 
An element $g \in Sp(2, R)$ defines the canonical transformation
\begin{equation}
q' = aq - bp, \qquad  p' = - cq + dp.
\end{equation}
and we see immediately that the time-of-arrival transforms according to eq.\ (\ref{ProTrasf}).

In the corresponding quantum system, one can define a (Hermitian, but not self-adjoint) time-of-arrival operator by writing, in the Heisenberg picture,
\begin{equation} \label{Time3}
T = - \frac 1 2 \left( Q P^{-1} + P^{-1} Q \right).
\end{equation}
where $P$ and $Q$ are the operators corresponding to the canonical variables $p$ and $q$. A complete treatment can be found, for instance, in refs.\ \cite{BGL2,Giannitrapani,AOPRU,MLP}, where a wide bibliography can be found.  As we have seen above, instead of the operator $T$ one can introduce a POVM $\tau$ on the time axis covariant with respect to the time translation group. It may be interesting to try to impose the covariance with respect to the larger group $Sp(2, R)$ of the linear canonical transformations. In this case, the unitary representation $U$ is called the {\it metaplectic} representation and it is double-valued \cite{MQ,GS}. In other words, $U$ is an unitary representation of the metaplectic group $Mp(2)$, which is a double covering of $Sp(2, R)$.
                             
In order to construct a covariance system, it is convenient \cite{Cattaneo,CH,Werner} to introduce an auxiliary {\it imprimitivity system} \cite{Mackey,Mackey2}. In Section 2 we classify the relevant imprimitivity systems and the corresponding unitary representations, which, according to Mackey's imprimitivity theorem, can be described as induced representations. In Section 3 we consider the Lie algebra of $\cal G$, we interpret one of the generators as the energy and we require that it is a positive operator. Then we recall the classification of the positive-energy representations of $SL(2, R)$ \cite{Bargmann}. In Section 4 we find the positive-energy representations contained in the induced representations found in Section 2. In Section 5 we use the results of the preceding Sections to find the most general covariance system. In particular, we show that the problem is soluble for every choice of the positive-energy representation $U$.

In Section 6 we rederive some of the results of Section 4 by means of a general theorem  \cite{Mackey,Mackey2} which generalizes the Frobenius reciprocity theorem. This powerful method is useful in the treatment of more difficult problems \cite{PT}, when the direct method used in Section 4 becomes too complicated. In Section 7 we introduce the generalizations necessary to treat the case in which $U$ is a ray representation of $\cal G$, namely a unitary representation of its universal covering $\tilde{\cal G}$. Then we treat the  model introduced above, described by the metaplectic representation, and we find that a covariant POVM exists only for states with negative parity.

\bigskip

\section{Imprimitivity systems.}  

An imprimitivity system \cite{Mackey,Mackey2} is a covariance system in which the POVM is a spectral, namely a normalized projection-valued, measure.  We use the notations introduced in the preceding Section and we consider the spectral measure $I \to E(I)$ on the space $\hat{\cal M}$ and a unitary representation $g \to V(g)$ of the group $\cal G$ in the Hilbert space ${\cal H}'$. If
\begin{equation} 
V(g) E(I) V(g)^{\dagger} = E(\Lambda(g)I),
\end{equation}
we have an imprimitivity system.

If $A: {\cal H} \to {\cal H}'$ is an intertwining operator between the representations $U$ and $V$, namely if
\begin{equation} 
A U(g) = V(g) A,
\end{equation}
one can easily see that
\begin{equation} 
I \to \tau(I) = A^+ E(I) A
\end{equation}
is a POVM which, together with the representation $U$, forms a covariance system, which is normalized if $A^+ A = 1$, namely if $A$ is isometric. It has been shown \cite{Cattaneo,CH,Werner} that all the covariance systems can be obtained in this way with a suitable choice of the imprimitivity system and of the intertwining operator.

Then, in order to find the required covariance systems, the first step is the construction of all the imprimitivity systems based on the group $\cal G$ and the homogeneous space $\hat{\cal M}$. This can be obtained by means of Mackey's imprimitivity theorem \cite{Mackey,Mackey2}, which gives a representation of the space ${\cal H}'$ by means of square integrable vector-valued functions $\phi(t)$ defined on $\hat{\cal M}$, on which $V$  acts as an induced representation. The Lebesgue measure $dt$ on $\cal M$ defines a measure on $\hat{\cal M}$ which is quasi-invariant under the action of $\cal G$ (the point at infinity has a vanishing measure). Then we can put
\begin{equation} 
\|\phi\|^2 = \int_{\hat{\cal M}} \|\phi(t)\|^2 \, dt.
\end{equation}
The projection operators $E(I)$ are given by 
\begin{equation} 
[E(I) \phi](t) = f_I(t) \phi (t),
\end{equation}
where $f_I(t)$ is the characteristic function of the set $I$.

From the physical point of view, the probability that the observable described by $\tau$, when measured on the state described by the normalized vector $\psi$, gives a result contained in the set $I$ is given by
\begin{equation} 
(\psi, \tau (I) \psi) = (A \psi, E(I) A \psi) = \int_I \rho(t) \, dt, 
\end{equation}
where the probability density $\rho(t)$ is given by
\begin{equation} \label{Prob}
\rho(t) = \|\phi(t)\|^2,  \qquad \phi = A \psi.
\end{equation}

In order to define the induced representation $V$, we have to choose a point of $\hat{\cal M}$, for instance $t_0 = 0$, and to consider the corresponding stabilizer subgroup defined by  $\Lambda(g) t_0 = t_0$, which, in the case we are considering, is just the subgroup $\cal S$ introduced in the preceding Section. For each point $t \in \hat{\cal M}$, we choose an element $g_t \in {\cal G}$ with the property $\Lambda(g_t) t_0 = t$. For instance, if $t$ is finite, we can choose
\begin{equation} 
g_t =  \left( \begin{array}{cc} 1 & t \\ 0 & 1 \end{array} \right).
\end{equation}
Then, the induced representation $V$ of $\cal G$, which depends on the unitary representation $S$ of $\cal S$, is given by 
\begin{equation} \label{Induced}
[V(g)\phi](t) = \left|\frac{dt'}{dt}\right|^{1/2} S(g_t^{-1} g g_{t'}) \phi(t'),
\end{equation}
where
\begin{equation} 
t' = \Lambda(g^{-1}) t = \frac{dt-b}{a - ct}, \qquad \frac{dt'}{dt} = (a - ct )^{-2}.
\end{equation}
The function $\phi$ takes its values in the Hilbert space of the representation $S$. 

For the elements of $\cal S$ we use the notation
\begin{equation}  \label{Sub}
s = (u, d) = \left( \begin{array}{cc} 1 & 0 \\ u & 1 \end{array} \right)
\left( \begin{array}{cc} d^{-1} & 0 \\ 0 & d \end{array} \right) =
\left( \begin{array}{cc} d^{-1} & 0 \\ 0 & d \end{array} \right)
\left( \begin{array}{cc} 1 & 0 \\ ud^{-2} & 1 \end{array} \right).
\end{equation}  
Since we have
\begin{equation} 
g_t^{-1} g g_{t'} = (c (a-ct)^{-1}, (a-ct)^{-1}),
\end{equation}                 
eq.\ (\ref{Induced}) can be written more explicitly in the form
\begin{equation} \label{Induced2}
[V(g)\phi](t) = |a-ct|^{-1} S(c (a-ct)^{-1}, (a-ct)^{-1}) \phi(t').
\end{equation} 

We remark that $\cal S$ is the product of the subgroup ${\cal S}_0$ which contains the elements with $d > 0$ and $\cal Z$ which contains the two elements $g = \pm 1$ (multiples of the identity matrix). One can easily see that ${\cal S}_0$ is connected and simply connected and that $\cal Z$ is the center of $\cal G$. The irreducible unitary representations (IURs) of $\cal S$ can be written in the form
\begin{equation} \label{RapS}
S(u, d) = S_0(u, |d|) d^{2 \nu} |d|^{-2 \nu}, \qquad \nu = 0, 1/2,
\end{equation}
where $S_0$ is an IUR of ${\cal S}_0$.     

We also see that
\begin{equation} 
(u, d)(u', d') = (u + d^2 u', dd'),
\end{equation}
namely ${\cal S}_0$ is a semi-direct product isomorphic to the one-dimensional affine group. In order to find its irreducible unitary representations, we use the classical procedure used by Wigner in his treatment of the IURs of the Poincar\'e group \cite{Wigner}. The representation $S_0$ operates on a space of square integrable functions $\eta(\kappa)$  defined on an orbit of the of the ``momentum space'' and the ``translation'' subgroup acts in the following way:
\begin{equation} 
[S_0(u, 1)\eta](\kappa) = \exp(-iu \kappa) \eta(\kappa).
\end{equation}
The action of the ``homogeneous'' subgroup, composed of the elements of the form $(0, d)$, on the ``momentum'' space is $\kappa \to d^{-2} \kappa$. The ``momentum space'' (a real line) is decomposed into three orbits:
\begin{equation} 
O_0 = \{0\}, \qquad O_+ = \{\kappa > 0\}, \qquad O_- = \{\kappa < 0\}.
\end{equation}

The IURs corresponding to the orbit $O_0$ do not depend on the variable $u$, are one-dimensional and have the form
\begin{equation} 
S_0^{\gamma}(u, |d|) = |d|^{-i\gamma},  \qquad -\infty < \gamma < +\infty.
\end{equation}
In the other two cases, in each orbit $O_{\sigma}$ with $\sigma = \pm 1$, we choose a representative element $\kappa = \sigma$. In both cases the stability subgroup contains only the unit element and the corrresponding IURs of ${\cal S}_0$ have the form  
\begin{equation} \label{RapS2}
[S_0^{\sigma}(u, |d|)\chi](\kappa)  = \exp(-iu \sigma \kappa) |d| \chi(d^2 \kappa), 
\end{equation}
where for $\sigma = -1$ we have changed the sign of $\kappa$ in such a way that we always have $\kappa > 0$.  The norm is given by
\begin{equation} 
\|\chi\|^2 = \int_0^{\infty} |\chi(\kappa)|^2 \, d\kappa.
\end{equation}

In conclusion, from eq.\ (\ref{Induced2}) we have two classes of induced representations of $\cal G$. The representations of first class, induced by the representations
\begin{equation} 
S^{\nu \gamma}(u, d) = S^{\gamma}_0(u, |d|) d^{2 \nu} |d|^{-2 \nu},
\end{equation}
can be written in the form
\begin{equation} \label{Classe1}
[V^{\nu \gamma}(g)\phi](t) = (a-ct)^{-2\nu} |a-ct|^{i \gamma + 2\nu -1} \phi(t'),
\end{equation}  
\begin{equation} 
\|\phi\|^2 = \int_{-\infty}^{+\infty} |\phi(t)|^2 \, dt.
\end{equation}
The representations of the second class are induced by the representations
\begin{equation} 
S^{\nu \sigma}(u, d) = S^{\sigma}_0(u, |d|) d^{2 \nu} |d|^{-2 \nu},
\end{equation}
and are described by
\begin{displaymath}
[V^{\nu \sigma}(g)\phi](\kappa, t) =
\end{displaymath}
\begin{equation} \label{Classe2}
= \exp(-i \sigma c (a-ct)^{-1} \kappa) (a-ct)^{-2\nu} |a-ct|^{2\nu - 2}  \phi(\kappa (a-ct)^{-2}, t'),
\end{equation}
\begin{equation} 
\|\phi\|^2 = \int_{-\infty}^{+\infty} dt \int_0^{+\infty} |\phi(\kappa, t)|^2 \, d\kappa.
\end{equation}

\bigskip

\section{Representations of the Lie algebra $sl(2, R)$.}  

The Lie algebra ${\cal L} = sl(2, R)$ of $SL(2, R)$ is composed of the real $2 \times 2$ matrices $h$ with the property
\begin{equation}
{\rm Tr}\, h = 0.
\end{equation}
We introduce a basis composed of the elements
\begin{equation}
e = \left( \begin{array}{cc}
0 & -1 \\ 0 & 0
\end{array} \right), \qquad
d = \frac 1 2 \left( \begin{array}{cc}
1 & 0 \\ 0 & -1
\end{array} \right), \qquad
c = \left( \begin{array}{cc}
0 & 0 \\ 1 & 0
\end{array} \right), 
\end{equation}
which represent, respectively, infinitesimal time translations, infinitesimal time dilatations and a third kind of infinitesimal projective transformations. We indicate by $-iE$, $-iD$ and $-iC$ the corresponding generators of a given unitary representation. They satisfy the commutation relations
\begin{equation}
[E, C] = -2i D, \qquad [D, E] = i E, \qquad [D, C] = -i C.
\end{equation}

If the unitary representation $U$ operates on physical states, the energy operator $E$ must be positive.  We also assume that the trivial representation of $\cal G$, which has a vanishing energy, is not contained in $U$; in fact, an invariant state cannot correspond to a normalizable probability distribution. We indicate by ${\cal L}_+$ the smallest closed convex cone in $\cal L$ invariant with respect to the adjoint representation which contains the element $e$. One can easily see that all the elements of this cone are represented by positive operators. From the relation
\begin{equation}
\exp\left(\frac \pi 2 (e+c)\right) e \exp\left(-\frac \pi 2 (e+c)\right) = c,
\end{equation}
we see that $c$ and $e + c$ belong to ${\cal L}_+$. It follows that the operator
\begin{equation}
K = \frac 1 2 (E + C)
\end{equation}
must be positive.

From eqs.\ (\ref{Induced2}) and (\ref{RapS}), we obtain
\begin{equation}
\exp(-2i \pi K) = V(\exp(\pi (e+c))) = V(-1) = (-1)^{- 2 \nu}, 
\end{equation}
and we see that $K - \nu$ must have integral eigenvalues. If we put
\begin{equation}
A_{\pm} = \frac {E - C} 2 \pm i D,
\end{equation}
we find
\begin{equation}
[K, A_{\pm}] = \pm A_{\pm}
\end{equation}
and we see that the operators $A_{\pm}$ play the role of rising and lowering operators. It follows that, for a given positive-energy IUR, the eigenvalues of $K$ are $k, k+1, k+2,\ldots$. 

A complete treatment of the IURs of $SL(2, R)$ is given in refs.\ \cite{Bargmann,GGV}, where it is shown that the positive-energy representations (together with the negative-energy ones) form the discrete series and are labelled by the index $k = 1/2, 1, 3/2,\ldots$. We indicate them by $D^k(g)$. The IURs of the universal covering of $SL(2, R)$ are treated in ref.\ \cite{Pukanszky,Sally}. The positive-energy IURs of the metaplectic group are labelled by the index $k = 1/4, 1/2, 3/4, 1,\ldots$. We see that the equivalence classes of positive-energy IURs of $SL(2, R)$ or of its double covering $Mp(2)$ form a countable set. It follows that the representation $U$ that acts on the ``physical'' states is a direct sum of IURs and the more subtle concept of direct integral is not needed for its treatment.

In the space of the positive-energy IUR  $D^k$ we can introduce a ``canonical'' basis $ Z^k_m\}$ with the properties \cite{Bargmann}
\begin{equation}
K Z^k_m = m Z^k_m, \qquad m = k, k + 1,\ldots,
\end{equation}
\begin{equation}
A_{\pm} Z^k_m = (m(m \pm 1) - k (k - 1))^{1/2} Z^k_{m \pm 1},
\end{equation}
and, in particular, we can characterize the vector $Z^k_k$ by means of the conditions
\begin{equation}  \label{Cond}
A_- Z^k_k = 0, \qquad K Z^k_k = k Z^k_k.
\end{equation}  
In the canonical basis the representation operators $D^k$ are described by the matrices
\begin{equation}
D^k_{m m'}(g) = (Z^k_m, D^k Z^k_{m'}).
\end{equation}  

\bigskip

\section{The positive-energy subrepresentations.}  

In order to treat the intertwining operator $A$, we have to find  which IURs $D^k$ are contained in the representations (\ref{Classe1}) and (\ref{Classe2}). We have already shown that $k - \nu$ must be integral. A direct approach to this problem is to find in the representation space a vector $Z^k_k$ which satisfies the conditions (\ref{Cond}). We recall that the generators of the infinitesimal transformations are defined on an invariant dense linear subspace  ${\cal D} \in {\cal H}'$. 

If we consider first a representation  of the kind (\ref{Classe1}) we obtain for the generators of the infinitesimal transformations
\begin{equation}  
E = i \frac{d}{dt},
\end{equation}
\begin{equation}  
D = - \frac{i + \gamma}{2} -it \frac{d}{dt},
\end{equation} 
\begin{equation}  
C = (i + \gamma) t + it^2 \frac{d}{dt},
\end{equation}  
\begin{equation}  
K = \frac{i + \gamma}{2} t + \frac{i}{2}(1 + t^2) \frac{d}{dt},
\end{equation}
\begin{equation}  
A_{\pm} = - \frac{i + \gamma}{2} (t \pm i) - \frac{i}{2}(t \pm i)^2 \frac{d}{dt}.
\end{equation}

From the conditions (\ref{Cond}) we obtain by means of simple calculations \begin{equation} \label{Solution} 
\phi(t) = \alpha (t - i)^{i\gamma - 1}, \qquad  k = \frac{1 - i \gamma}{2}.
\end{equation}
Since $k - \nu$ must be integral, we have to put $\gamma = 0$ and $\nu = 1/2$. In conclusion, we have shown that the representation $V^{\nu \gamma}$ contains a positive-energy subrepresentation only if $\gamma = 0$ and $\nu = 1/2$ and this subrepresentation is $D^{1/2}$.  

The canonical basis for this representation can be obtained starting from eq.\  (\ref{Solution}), which in the interesting case, after normalization, takes the form 
\begin{equation} 
Z^{1/2}_{1/2}(t) = \pi^{- 1/2} (1 + it)^{- 1}.
\end{equation} 
By successive applications of the raising operator $A_+$ we obtain 
\begin{equation} 
Z^{1/2}_m(t) = \pi^{- 1/2} (1 - it)^{m - 1/2} (1 + it)^{- m - 1/2}.
\end{equation}  

Then we consider a representation  of the kind (\ref{Classe2}). The generators of the infinitesimal transformations are given by
\begin{equation}  
E = i \frac{\partial}{\partial t},
\end{equation} 
\begin{equation}  
D = - i - i \kappa \frac{\partial}{\partial \kappa} -it \frac{\partial}{\partial t},
\end{equation}
\begin{equation}  
C = \sigma \kappa + 2it + 2it \kappa \frac{\partial}{\partial \kappa} + it^2 \frac{\partial}{\partial t},
\end{equation}  
\begin{equation}  
K = \frac{1}{2} \sigma \kappa + i t +  i t \kappa \frac{\partial}{\partial \kappa} + \frac{i}{2}(1 + t^2) \frac{\partial}{\partial t},
\end{equation}
\begin{equation}  
A_{\pm} = -\frac{1}{2} \sigma \kappa - i(t \pm i)\left(1 + \kappa \frac{\partial}{\partial \kappa}\right) - \frac{i}{2}(t \pm i)^2 \frac{\partial}{\partial t}.
\end{equation}

The conditions (\ref{Cond}) have the solution
\begin{equation} \label{Solution2} 
\phi(\kappa, t) = \alpha (1 + it)^{-2k} \kappa^{k - 1} \exp\left(\frac{-\sigma \kappa}{1 + it}\right).
\end{equation}
We see that this function can be square integrable only if $\sigma = 1$ and $k > 1/2$. In conclusion, we have shown that the representation $V^{\nu \sigma}$ contains positive-energy subrepresentations only if $\sigma = 1$ and the maximal positive-energy subrepresentation is $D^1 \oplus D^2 \oplus \ldots$ if $\nu = 0$ and $D^{3/2} \oplus D^{5/2} \oplus \ldots$ if $\nu = 1/2$.

The first function $Z^k_k$ of the canonical basis for the representation $D^k$ can be obtained by normalizing eq.\  (\ref{Solution2}). The other functions are given by successive applications of the raising operator $A_+$. In this way we obtain 
\begin{displaymath} 
Z^k_m(\kappa, t) =  \left( \frac{2(2k - 1) (m - k)!}{\pi (m + k - 1)!}\right)^{1/2}
\end{displaymath} 
\begin{equation} 
\left(\frac{1 - it}{1 + it}\right)^m \frac{1}{1 + t^2} \, x^{k - 1} L^{(2k - 1)}_{m - k}(x) \exp\left(\frac{-\kappa}{1 + it}\right),
\end{equation} 
where
\begin{equation} 
x = \frac{2\kappa}{1 + t^2} = 2 {\bf Re} \left(\frac{\kappa}{1 + it}\right).
\end{equation} 
We have used the following properties of the Laguerre polynomials \cite{AS}
\begin{equation} 
L^{(\alpha)}_0(x) = 1, \qquad (n + 1) L^{(\alpha)}_{n + 1}(x) = x \frac{d}{dx} L^{(\alpha)}_n(x) + (n + 1 +\alpha - x) L^{(\alpha)}_n(x). 
\end{equation}

\bigskip

\section{The covariance systems.}  

As we have already observed, the unitary representation $U$ can be decomposed into the direct sum of IURs of the kind $D^k$. If in every invariant subspace we introduce a canonical basis, we can describe the state vector $\psi \in {\cal H}$ by means of the coefficients $\psi_{\alpha k m}$ and the representation $U$ takes the form
\begin{equation}
[U(g)\psi]_{\alpha k m} = \sum_{m'} D^k_{mm'}(g) \psi_{\alpha k m'}.
\end{equation}

The intertwining operator $A$ transforms the vector $\psi$ into a positive-energy vector $A\psi \in {\cal H}'$. The representation $V$ acting on ${\cal H}'$ can be decomposed into the direct sum of representation of the kind $V^{\nu \gamma}$ or $V^{\nu \sigma}$, which contain positive-energy subrepresentations of the kind $D^k$. If we introduce in the corresponding invariant subspaces the canonical bases introduced in Section 4, we can write 
\begin{equation}
A\psi = \sum_{\beta k m} \phi_{\beta k m} Z^k_{\beta m}, 
\end{equation}                                
where the index $\beta$ labels the subspaces in which equivalent representations $V^{\nu \gamma}$ or $V^{\nu \sigma}$ operate.

It follows from the Schur lemma that the matrix that represents the intertwining operator $A$ is diagonal in the indices $k, m$ and does not depend on the value of $m$, namely we have 
\begin{equation}
\phi_{\beta k m} = \sum_{\alpha} A^k_{\beta \alpha} \psi_{\alpha k m},
\end{equation} 
\begin{equation}
A\psi = \sum_{\alpha \beta k m} A^k_{\beta \alpha} \psi_{\alpha k m} Z^k_{\beta m}. 
\end{equation} 
Since $A$ is isometric, we have
\begin{equation}
\sum_{\beta} \overline{A^k_{\beta \alpha}} A^k_{\beta \alpha'} = \delta_{\alpha \alpha'}.
\end{equation} 

The probability density (\ref{Prob}) is given by
\begin{displaymath} 
\rho(t) = \sum_{\alpha} | \sum_m \psi_{\alpha, 1/2, m} Z^{1/2}_{m}(t) |^2 + 
\end{displaymath} 
\begin{equation}
+ \sum_{\nu = 0, 1/2} \sum_{\beta} \int | \sum_{k=\nu+1}^{\infty} \sum_{\alpha m} A^k_{\beta \alpha} \psi_{\alpha k m} Z^k_m (\kappa, t) |^2 \, d\kappa. 
\end{equation}
Note that there is no interference between terms with integral and half-odd values of $k$ and between terms with $k = 1/2$ and other values of $k$.

The integration over the variable $\kappa$ can be performed, since we have 
\begin{equation}
\int_0^{\infty}  Z^k_m(\kappa, t) \overline{Z^{k'}_{m'}(\kappa, t)}  \, d\kappa = \pi^{-1} \left(\frac{1 - it}{1 + it}\right)^{m-m'} \frac{1}{1 + t^2} C^{k k'}_{m m'}.
\end{equation} 
The coefficients $C^{k k'}_{m m'}$ are given by an integral containing the product of two Laguerre polynomials, which can be expressed in terms of a generalized hypergeometric series \cite{PBM}:
\begin{displaymath}
C^{k k'}_{m m'} = \left(\frac{(2k - 1) (2k' - 1) (m + k - 1)!}{(m - k)! (m' - k')! (m' + k' - 1)!}\right)^{1/2} 
(k' - k + 1)_{m' - k'}\frac{(k + k' - 2)!}{(2k - 1)!}
\end{displaymath}
\begin{equation}
{}_3F_2(k - m, k + k' - 1, k - k'; 2k, k - m'; 1).
\end{equation}
For some values of the parameters the generalized hypergeometric series can be summed and we obtain
\begin{equation}
C^{k k'}_{m m} = \delta _{k k'},
\end{equation}
which ensures the orthogonality of the functions $Z^k_m$, and
\begin{equation}
C^{k k}_{m m'} =  C^{k k}_{m' m} =\left(\frac{(m + k - 1)! (m' - k)!}{(m - k)! (m' + k - 1)!}\right)^{1/2}, \qquad m' \geq m.
\end{equation}
 
When the representation $U$ is irreducible, namely $U = D^k$, the matrices $A^k_{\beta \alpha}$ disappear and we obtain the simple formula 
\begin{equation} \label{Simple}
\rho(t) = \pi^{-1} \frac{1}{1 + t^2} \sum_{m m'} \left(\frac{1 - it}{1 + it}\right)^{m-m'} C^{k k}_{m m'} \psi_{k m} \overline{\psi_{k m'}}.
\end{equation}
This formula is valid also for $k = 1/2$, when we have $C^{1/2, 1/2}_{m m'} = 1$. 

\bigskip

\section{Use of the Frobenius reciprocity theorem.}  

The results of Section 4 can be interpreted and partially rederived by means of a generalization due to Mackey \cite{Mackey,Mackey2} of the Frobenius reciprocity theorem.  

We recall that a unitary representation is called {\it square integrable} if its matrix elements are square integrable functions on the group. The square integrable IURs are contained as subrepresentations in the regular representation and they have a non-vanishing Plancherel measure.  For our purposes, it is sufficient to use the following consequence of the Mackey reciprocity theorem. We assume that all the groups considered are locally compact and separable. 	

\noindent {\bf Theorem:} Let $D$ be a square integrable IUR of $\cal G$ and $S$ an IUR of the closed subgroup ${\cal S} \subset {\cal G}$. We indicate by ${\rm ind}\,S$ the representation of $\cal G$ induced by $S$ and by $D|_{\cal S}$ restriction of $D$ to $\cal S$. We assume that ${\rm ind}\, S$ contains $D$ as a subrepresentation with multiplicity $n$ (possibly zero or infinity) and that $D|_{\cal S}$ contains $S$ as a subrepresentation with multiplicity $n'$ (possibly zero or infinity). Then, if $D$ is square integrable and $n' > 0$, we have $n = n'$ and $S$ is square integrable. Symmetrically, if $S$ is square integrable and $n > 0$, we have $n' = n$ and $D$ is square integrable.

It is well-known \cite{Bargmann} that the IURs $D^k$ are square-integrable for $k > 1/2$, but not for $k = 1/2$.  With the notation introduced in Section 2 we have  
\begin{equation}
{\rm ind}\,S^{\nu \gamma} = V^{\nu \gamma}, 	\qquad
{\rm ind}\,S^{\nu \sigma} = V^{\nu \sigma}
\end{equation}   
and it is easy to show that the IURs $S^{\nu \sigma}$ are square integrable, while the IURs $S^{\nu \gamma}$ have not this property.
  
We also need to consider the restriction $D^k|_{\cal S}$.  In ref.\ \cite{GGV} the representations $D^k$ are defined in a space of functions of the complex variable $z$ analytic in the lower half plane by means of the formula 
\begin{equation}
[D^k(g)\eta](z) = (bz + d)^{2k - 2} \eta\left(\frac{az + c}{bz + d}\right),
\end{equation}  
with a suitable definition of the norm.  Its restriction to $\cal S$, with the notation of eq.\ (\ref{Sub}) is given by 
\begin{equation}
[D^k(u,d)\eta](z) = d^{2k - 2} \eta(d^{-2}(z + u)).
\end{equation}   
We introduce the new function 
\begin{equation}
\chi(\kappa) = \kappa^{k - 1/2} \int_{-\infty}^{\infty} \exp(ikz) \eta(z) \, dz.
\end{equation}   
Since $\eta$ is analytic in the lower half plane (and has a suitable behaviour at infinity), this integral vanishes for $\kappa < 0$. The transformation property takes the form 
\begin{equation}
[D^k(u,d)\chi](\kappa) = d^{2k} |d|^{1 - 2k} \exp(-i \kappa u) \chi(d^2 \kappa).
\end{equation} 
A comparison with eqs.\ (\ref{RapS}) and (\ref{RapS2}) shows that
\begin{equation}
D^k|_{\cal S} = S^{\nu, 1}, 
\end{equation}   
where $k - \nu$ is integral.

We deduce, in agreement with the results of Section 4, that $V^{\nu, -1}$  contains no representation of the kind $D^k$, while  $V^{\nu, 1}$ contains with multiplicity one just the representations $D^k$ with $k > 1/2$ and $k - \nu$ integral. No result concerning the representations $V^{\nu \gamma}$ can be obtained in this way. 

\bigskip

\section{The free particle model and projective representations.}  

In order to treat the model suggested in the Introduction, based on a free particle in one dimension,  we note that the generators of the metaplectic representation are given by
\begin{equation}
E = \frac 1 2 P^2, \qquad C = \frac 1 2 Q^2 , \qquad D = \frac 1 4 (QP + PQ), \qquad  K = \frac 1 4 (P^2 + Q^2).
\end{equation}
These fomulas substituted into eq.\ (\ref{Time2}) give eq.\ (\ref{Time3}). The operator $K$ is half the hamiltonian of an armonic oscillator with $\omega = m = \hbar = 1$ and its eigenvalues are $1/4, 3/4, 5/4,\ldots$. It follows that the metaplectic representation is reducible and is given by $D^{1/4} \oplus D^{3/4}$. Since this is a projective representation of ${\cal G} = SL(2, R)$ and a representation of its two-fold covering $Mp(2)$ or, more in general, of its universal covering $\tilde{\cal G}$, we have to adapt the methods introduced in the preceding Sections.

We indicate by $\tilde{\cal S}$ the inverse image of $\cal S$ under the covering mapping $\tilde{\cal G} \to {\cal G}$ and we see that $\hat{\cal M} = \tilde{\cal G} / \tilde{\cal S}$.  The connected component of the identity $\tilde{\cal S}_0$ is isomorphic to ${\cal S}_0$, which is simply connected; we identify these two groups and we indicate their elements with the symbol $(u, d)$ ($d > 0$) defined by eq.\ (\ref{Sub}). It follows that $\tilde{\cal S}$ is the product of $\tilde{\cal S}_0$ and the center $\tilde{\cal Z}$ of $\tilde{\cal G}$, which is composed of the elements of the form $z_n = \exp(\pi n (e + c))$, where $n$ is an integer and $\exp$ is the exponential mapping of $\tilde{\cal G}$ (which is not a group of matrices).
         
The IURs of $\tilde{\cal S}$ have the form 
\begin{equation} 
S((u, d)z_n) = S_0(u, d) \exp(2i \pi n \nu), \qquad 0 \leq \nu < 1.
\end{equation}
and, as in Section 2, we define the corresponding induced representations $V^{\nu, \gamma}$ and $V^{\nu, \sigma}$ by means of eq.\ (\ref{Induced}). It follows that
\begin{equation}
\exp(-2i \pi n K) = V(z_n) = \exp(2i \pi n \nu) 
\end{equation}
and we find also in this general case that $K - \nu$ has integral eigenvalues.  

The generators of the infinitesimal transformations are represented by the same differential operators found in Section 4, though they are defined in different linear subspaces $\cal D$.  From the conditions (\ref{Cond}) we obtain again the solutions  (\ref{Solution}) and (\ref{Solution2}). The first is acceptable only if $\gamma = 0$ and $\nu = 1/2$ and the second is square integrable only if $\sigma = 1$ and $k > 1/2$.  It follows that $V^{\nu, 1}$ contains the positive-energy IURs $D^k$ with $k = \nu, \nu + 1,\ldots$ if $1/2 < \nu < 1$ and $k = \nu + 1, \nu + 2,\ldots$ if  $0 \leq \nu \leq 1/2$.

We see that $D^{3/4}$ is contained in $V^{1, 3/4}$, but $D^{1/4}$ is not contained in any of the induced representations.  It follows that a normalized POVM can be found only for free particle states which belong to the space of the representation $D^{3/4}$, which is spanned by the odd eigenstates of the harmonic oscillator hamiltonian $K$. In other words, we have to require that the wave function $\psi(q)$ of the particle has odd parity and therefore it vanishes for $q = 0$. A possible physical interpretation is to assume that there is an infinite potential barrier at $q = 0$ and $t$ is the time at which the particle is reflected by the barrier.
 
We introduce in $\cal H$ a canonical basis composed of eigenfunctions of the operator $K$ and we indicate by $\psi_m$ (m = 3/4, 7/4,\ldots) the coefficients of the corresponding expansion of $\psi \in {\cal H}$. By reasoning as in Section 5 we obtain eq.\ (\ref{Simple}) with $k = 3/4$ and the covariant POVM is uniquely determined. The wave function of the particle is given by
\begin{displaymath} 
\psi(q) = \pi^{-1/4} \exp\left(- \frac{q^2}{2}\right) \sum_{m} \psi_m (2^n n!)^{-1/2} H_n(q),  
\end{displaymath}
\begin{equation} 
\qquad n = 2m - 1/2 = 1, 3,\ldots. 
\end{equation}

If we require only the covariance with respect to the time translations, the POVM is not unique any more. In refs.\ \cite{BGL2,Giannitrapani} a very natural choice of the POVM has been suggested, which, for wave functions with a given parity,  leads to the following probability density:  
\begin{equation} \label{Simple1} 
\rho(t) = \pi^{-1} \left| \int_0^{\infty} \exp\left(\frac{i}{2} tp^2 \right) \tilde\psi(p) p^{1/2} \, dp \right|^2.
\end{equation}
If only the amplitude with $m = k = 3/4$ is present, we obtain
\begin{equation} 
\rho(t) =  \left(\frac 2 \pi \right)^{3/2}
\left(\Gamma\left(\frac 5 4 \right)\right)^2  \left(\frac{1}{1 + t^2}\right)^{5/4},
\end{equation}
while eq. (\ref{Simple}) gives
\begin{equation} 
\rho(t) = \pi^{-1} \frac{1}{1 + t^2}.
\end{equation}
We see that the POVM described by eq.\ (\ref{Simple1}) does not coincide with the one defined by eq.\ (\ref{Simple1}) and therefore it cannot be covariant with respect to the linear canonical transformations.

\end{document}